\RequirePackage[bookmarksnumbered,unicode]{hyperref}
\documentclass[sigconf,nonacm]{acmart}
\usepackage{tikz}
\AtBeginDocument{%
  \providecommand\BibTeX{{%
    \normalfont B\kern-0.5em{\scshape i\kern-0.25em b}\kern-0.8em\TeX}}}

\makeatletter
\def\@ACM@checkaffil{
    \if@ACM@instpresent\else
    \ClassWarningNoLine{\@classname}{No institution present for an affiliation}%
    \fi
    \if@ACM@citypresent\else
    \ClassWarningNoLine{\@classname}{No city present for an affiliation}%
    \fi
    \if@ACM@countrypresent\else
        \ClassWarningNoLine{\@classname}{No country present for an affiliation}%
    \fi
}
\makeatother

\begin{document}

\title{S3C2 Summit 2025-09: \\ Industry Secure Supply Chain Summit}

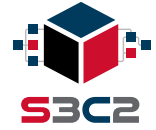
\author{Md Atiqur Rahman$^{*}$, Yasemin Acar$^{\dagger}$, Michel Cukier$^{\ddagger}$, William Enck$^{*}$,\\Christian Kästner$^{\mathsection}$, Alexandros Kapravelos$^{*}$, Dominik Wermke$^{*}$, Laurie Williams$^{*}$}

\def \authors{Elizabeth Lin, Jonah Ghebremichael, William Enck, Yasemin Acar, Michel Cukier, Alexandros Kapravelos, Christian Kästner, Laurie Williams}

\affiliation{%
    \institution{ $^*$North Carolina State University, Raleigh, NC, USA}
}
\affiliation{%
    \institution{$^\dagger$Paderborn University, Paderborn, Germany and George Washington University, DC, USA}
}
\affiliation{%
    \institution{$^\ddagger$University of Maryland, College Park, MD, USA}
}
\affiliation{%
    \institution{ $^\mathsection$Carnegie Mellon University, Pittsburgh, PA, USA}
}

\renewcommand{\shortauthors}{Secure Software Supply Chain Center (S3C2)}
\renewcommand{\shorttitle}{S3C2 Summit 2025-09: Secure Supply Chain Summit (West Coast)}

\begin{abstract}

Today's digital ecosystem relies heavily on software supply chains, which enable developers to reuse code and ship software at scale.
However, a single vulnerable component can jeopardize the entire supply chain.
In recent years, cyberattacks in software supply chains have become increasingly common.
These attacks can disrupt critical systems and put organizations, including major software companies, government agencies, and open-source contributors, at risk.
This growing threat has led to increased attention from both the software industry and the U.S.\ government toward strengthening software supply chain security.

On September 15, 2025, three researchers from the NSF-backed Secure Software Supply Chain Center (S3C2) convened a Secure Software Supply Chain Summit, bringing together 10 practitioners from 8 organizations across diverse domains.
The goals of the Summit were threefold: (1) to facilitate cross-industry sharing of practical experiences and challenges in securing software supply chains; (2) to foster new collaborations among participants; and (3) to identify pressing challenges to guide future research directions. The Summit featured discussions on six central topics: vulnerable dependencies, component and container choice, malicious commits, build infrastructure, culture, and the role of LLMs in the supply chain.
For each topic, participants engaged with a curated set of discussion questions designed to gather insights and pain points.

This report summarizes the key takeaways from these discussions. Each section highlights which topics continued from previous summits and which ideas emerged for the first time in this summit; the full list of initial discussion prompts is provided in the appendix.
\end{abstract}




\maketitle

\begin{tikzpicture}[overlay, remember picture]
\node[anchor=north west, 
      xshift=17.5cm, 
      yshift=-2.1cm] 
     at (current page.north west) 
     {\includegraphics[width=2.1cm]{S3C2_logo.pdf}}; 
\end{tikzpicture}

\section{Introduction}
\label{sec:intro}

Building on S3C2's ongoing efforts to strengthen the nation's software supply chain, a day-long Secure Software Supply Chain Summit was held on September 15, 2025.
Organized by three researchers from the NSF-backed Secure Software Supply Chain Center (S3C2)\footnote{https://s3c2.org/}, the event brought together 10 practitioners from 8 companies for focused discussions on emerging challenges and best practices in software supply chain security.
The goals of the Summit were threefold:
(1) to facilitate cross-industry sharing of practical experiences and challenges in software supply chain security;
(2) to foster new collaborations among practitioners and researchers; and
(3) to identify key challenges and research needs to inform future directions.

The Summit was conducted under the Chatham House Rule, allowing participants to freely use information from the discussions while prohibiting attribution of remarks or disclosure of attendees' affiliations.
Accordingly, this report also adheres to that rule, and participant identities and organizational affiliations are not disclosed.
Ten participants were drawn from eight companies actively engaged in software security, with attendance intentionally limited to encourage open and candid dialogue among key stakeholders.

Discussion topics were selected through a pre-summit voting process to discuss high-relevance issues to the organizations represented.
The final set of topics included: vulnerable dependencies (Sec.~\ref{sec:vuln-dependencies}), component and container choice (Sec.~\ref{sec:choices}), malicious commits (Sec.~\ref{sec:malicious-commits}), build infrastructure (Sec.~\ref{sec:build-infrastructure}), security culture (Sec.~\ref{sec:culture}), and the role of large language models (LLMs) in the software supply chain (Sec.~\ref{sec:llm-supply-chain}).
Each session was moderated by an S3C2 researcher and began with a set of guiding questions designed to spark discussion; these guiding questions are provided in the appendix.
For each topic, this report summarizes discussion on recurring themes from prior summits~\cite{Summit1, Summit2, Summit3, Summit4, Summit5, Summit6, Summit7} and highlights new insights and challenges that emerged in this year's discussions.

Three S3C2 researchers (two professors and one Ph.D.\ student) recorded detailed notes during the sessions. The Ph.D.\ student prepared an initial draft of this report based on those notes, which was subsequently reviewed and revised by the professors.
The final version was reviewed by all Summit participants and additional S3C2 researchers with expertise in software supply chain security.

The following sections summarize the discussions and key takeaways from the September 2025 Secure Software Supply Chain Summit.

\section{Vulnerable Dependencies}
\label{sec:vuln-dependencies}
Managing vulnerable dependencies remains one of the most persistent challenges discussed across S3C2 summits.
Participants reiterated familiar difficulties, such as excessive alerts, inconsistent metadata, difficult choices about what to fix first, and constant churn from transitive dependencies and rapidly changing ecosystems.
At the same time, this summit raised new questions about trust, IDE-level exposure, agentic tooling, and how AI-mediated development is reshaping the concept of \emph{vulnerability} itself.

\subsection{Continued Themes}

The discussions revisited familiar challenges around dependency alerts and risk, trust and provenance gaps, and the limits of automation.

\paragraph{Vulnerability volume and prioritization fatigue}
Practitioners described being overwhelmed by a constant stream of dependency alerts from software composition analysis (SCA) tools and automated patching systems.
Common challenges included too many alerts, mixed signals from different scanners, and the need to decide which issues actually matter.

\paragraph{Risk beyond CVEs}
Participants also noted that CVEs alone are not enough to measure dependency risk. Signals such as maintainer activity, package update frequency, package age, the presence of obfuscated code, and end-of-life status were also mentioned as relevant.

\paragraph{Staying current at scale}
Another recurring challenge is to keep dependencies up to date at scale. Participants pointed to breaking changes, weak test coverage, and pressure to ship new releases.

\paragraph{Trust and provenance gaps}
As at earlier summits, participants mentioned that they are often unsure how much to trust upstream maintainers, vulnerability databases, or third-party-curated lists. There is still a strong need for more reliable indicators about a software's origin and who is responsible for it.

\paragraph{Automation limits}
Automation helps, but it cannot replace human judgment. Participants noted that automated update tools often create extra noise, break builds, or require developers to step in and validate changes.

\subsection{New Ideas}

The vulnerable dependencies topic session discussed new ideas around where risk shows up, how developers interact with it, and who should help manage trust.

\paragraph{Inner-loop dependency signals}
A new direction discussed was to show dependency risk directly in IDEs and pull request workflows.
Tools such as socket.dev and advisory-aware MCP integrations aim to help developers make decisions and triage issues closer to where code is written.

\paragraph{Developer endpoints as supply-chain attack surfaces}
This summit also discussed IDEs, plugins, and developer machines as part of the supply chain.
Concerns around MCP servers, internet-retrieved tools, and compromised endpoints drove discussion of ephemeral, cloud-based development environments.

\paragraph{Vetted dependency sets}
Participants expressed interest in having major platforms such as Google or GitHub publish vetted and audited sets of open-source libraries.
This would allow smaller teams to adopt these libraries with more confidence and reduce the burden of evaluating dependencies on their own.

\section{Component and Container Choice}\label{sec:choices}

Component and container choice continue to reflect a loose, developer-driven process in most organizations.
Teams primarily rely on tools like Dependabot to flag problems after a component is already in use, rather than evaluating them up front.
Some teams are starting to define more explicit rules for open-source scanning, license checks, and container baselines.

\subsection{Continued Themes}

Discussions about component and container choices returned to familiar patterns in how teams select, monitor, and update their dependencies.

\paragraph{Reactive dependency management}
As in previous summits, teams continue to rely heavily on SCA tools, mainly Dependabot, to identify issues.
Developers make most of the selection decisions.
Security teams step in only when SCA tools indicate concerns.

\paragraph{Trusted base images and registry scanning}
Organizations continued to build on well-known base container images and run regular vulnerability scans.
This year's panel again stressed the importance of scanning within artifact registries and internal repositories rather than relying on one-off or pull-request checks.
The main focus is on keeping container images up to date, well-monitored, and built on trusted foundations.

\subsection{New Ideas}
This section highlights new ways teams are thinking about dependency, build, and development environment risk.

\paragraph{Network-isolated CI/CD builds}
This year, network-isolated builds were discussed.
For network-isolated builds, after the dependencies are downloaded, build jobs should no longer have outbound network access.
Participants described this as a more reliable safeguard than relying only on reputation scores or policy-based controls.

\paragraph{Endpoint and IDE exposure as supply-chain risk}
The panel also highlighted risks that originate on developer machines, such as malicious plugins, misnamed packages, and shadow packages.
Some organizations are adding scanning directly in IDEs to catch suspicious packages before they reach CI.

\paragraph{Ecosystem-wide package-trust platforms}
Another new idea discussed was to bring together ecosystem signals such as maintainer history, release cadence, popularity, and vulnerability trends into a single package trust view. This platform could support LLM-based tools like Copilot to avoid suggesting risky dependencies.

\paragraph{GitHub Actions as supply-chain artifacts}

The panel emphasized treating GitHub Actions more like other third-party dependencies.
Teams described concrete practices such as pinning external Actions to specific commits, requiring approvals, and avoiding blind trust in the Actions marketplace.
Some teams are also moving toward immutable actions to reduce unexpected changes over time.

\paragraph{Registry-embedded container scanning}
Some organizations now run vulnerability scans directly within their artifact registries, rather than in CI pipelines.
This approach reduces friction during pull requests.
It also ensures that container images are rescanned as scanning tools improve.

\paragraph{Uncertainty around GitHub-hosted VM images}
A newer concern raised was the lack of visibility into GitHub-maintained runner images.
It is not always possible to know how often they are updated and which tools and language versions they include.
As a result, these VM images are increasingly treated as dependencies that also need closer review.

Together, these ideas reflect a push to reduce trust in default environments, catch risk earlier, and treat more parts of the build and development process as dependencies that need scrutiny.

\section{Malicious Commits}
\label{sec:malicious-commits}

Malicious commits remain a challenging supply chain risk to address.
Detecting these cases still relies on a mix of human judgment and imperfect heuristics.
The panel revisited long-standing concerns and shared new perspectives on contributor behavior, toolchain compromise, and signals that may indicate malicious intent.

\subsection{Continued Themes}

The discussion revisited long-standing challenges in understanding and defending against malicious commits.

\paragraph{Commit intent remains opaque}
As in earlier summits, participants noted that it is very challenging to determine intent from a single commit.
A potentially vulnerable or harmful change can appear to be just an honest mistake.
Intent might need to be viewed alongside the contributor's past context and downstream effects.

\paragraph{Trust boundaries and multi-party review}
The panel revisited familiar concerns around trust, including zero-trust assumptions and two-person reviews of pull requests.
Participants pointed to the risks of relying on a single maintainer or reviewer.
Social engineering also remains a significant risk factor from their perspective.

\paragraph{Process-level countermeasures}
The discussion then returned to practical steps such as sandboxing, clearer contribution policies, structured review gates, and tracking provenance as ways to reduce risk.
Participants noted that these safeguards can be hard to scale, but they still form the foundation of many security efforts.

\paragraph{Upstream uncertainty}
Consistent with previous years, attendees reiterated that upstream projects must be treated as potentially untrusted.
Malicious changes introduced in external dependencies can propagate silently.
This makes upstream vetting an unavoidable part of the problem.

These discussions highlight why malicious commits remain hard to detect and contain.

\subsection{New Ideas}

This section highlights new ways participants are trying to spot malicious behavior that may not be obvious from code changes alone.

\paragraph{Behavioral and identity signals}
A new focus this year was on contributor-centric heuristics.
Examples included unusual commit times, unexpected file modification patterns, or sudden changes to SSH/PGP keys.
Participants viewed these signals as lightweight hints that can point reviewers to changes that need more attention.

\paragraph{Tool-chain compromise paths}
Participants also discussed newer attack paths, such as compromised installers, modified build systems, and DLL sideloading, which can introduce malicious behavior without requiring changes to the source code.
Earlier summits focused mainly on commits in repositories, making this expansion in scope notably new.

\paragraph{Capability-change diffs}
Another new idea discussed was to examine changes in a package's capabilities, such as new file-system access, processes, or network behavior.
These changes can be considered as a signal during the review process.
The idea focuses on what a change allows the software to do, rather than just what the code looks like.

\section{Build Infrastructure}
\label{sec:build-infrastructure}

Build infrastructure security remains a complex and evolving problem.
Participants noted that modern CI/CD pipelines are often complex, highly automated, and composed of many interconnected components.
At the same time, these systems often lack clear boundaries of identity and reliable information about the origins of artifacts.
While frameworks like the Supply-chain Levels for Software Artifacts (SLSA) provide helpful guidance, many organizations still struggle with day-to-day implementation, scaling verification, and securing a wide range of runners and artifact flows.

\subsection{Continued Themes}

The discussion focused on recurring problems in build pipelines, runner environments, and provenance systems.

\paragraph{Pipeline fragility and trust boundaries}
Securing pipelines requires understanding trust relationships across the entire CI/CD system.
Participants noted that overly permissive runners, weak publishing controls, or piecemeal automation can undermine an otherwise well-secured pipeline.

\paragraph{SLSA adoption difficulties}
Many participants said they are still unclear about how to interpret SLSA levels and apply them in practice.
Organizations described difficulties aligning their internal systems with SLSA requirements, defining identity and access boundaries, and explaining expectations to developers.

\paragraph{Self-hosted vs.\ cloud runners}
The discussion then returned to an ongoing debate about where runners should live.
Some participants see cloud-hosted runners as safer because they are more isolated and consistently maintained.
Others prefer self-hosted runners for the visibility and control they provide.

\paragraph{Limited practical use of reproducible builds}
Reproducible builds have resurfaced as a goal that many teams would like to achieve, but few can use consistently in practice.
Participants highlighted some obstacles, including non-deterministic pipelines and differences in toolchains.
They also agreed that reproducibility alone is not sufficient to provide strong security guarantees.

\paragraph{Provenance and key-management challenges}
Verifying provenance at scale remains challenging for many teams.
Participants highlighted recurring issues, including the large volume of in-toto and SLSA metadata, key rotation overhead, verification across multiple systems, and the long lifespan of stored artifacts.

These challenges highlight why securing build systems remains difficult in practice.

\subsection{New Ideas}

Beyond these long-standing challenges, participants shared new ideas to strengthen trust and visibility in build systems.

\paragraph{Workload identity as a central gap}
When it is unclear which workload actually ran a build, it becomes hard to trust the resulting artifacts.
A significant new theme this year was the recognition that knowing which workload actually ran a build is required for trustworthy provenance.
Participants highlighted issues with inconsistent OIDC behavior and unclear identity models.
They also emphasized the importance of artifact registries in verifying identity before accepting uploads.

\paragraph{Applying SLSA to AI and agent pipelines}
Participants discussed how build provenance ideas could apply to AI and machine learning workflows, including agent-based systems.
Some ideas included treating agents as first-class identities and generating attestations for individual training steps.
Participants also discussed aligning common ML platforms, such as HuggingFace and internal model gateways, with existing provenance frameworks.

\paragraph{Parallel reproducible builds for anomaly detection}
Participants this year suggested using reproducible builds for more than just ensuring determinism.
By running builds in parallel and comparing results, teams can spot misconfigurations or signs of compromise.
Differences between builds could reveal problems or malicious interference in the build environment itself.

\section{Culture}
\label{sec:culture}

Culture emerged as a key factor in how organizations approach software supply chain security.
Participants emphasized that technical controls are only adequate when teams also have the right norms, incentives, and workflows to support secure behavior.
Much of the discussion centered on how leadership, developer experience, and day-to-day pressures influence whether security practices are actually implemented and sustained.

\subsection{Continued Themes}

The discussion revisited long-standing cultural challenges in the adoption and sustainment of security.

\paragraph{Top-down mandates and bottom-up adoption}
As in earlier summits, participants noted that cultural change needs both leadership support and buy-in from developers.
Leaders can set expectations; however, lasting adoption occurs only when developers believe that security (culture) is supported and worthwhile.

\paragraph{Minimizing developer friction}
Another familiar theme came up: developers are more likely to adopt security practices when they fit naturally into their existing workflows.
Participants repeated earlier calls for \emph{developer-first} approaches, including paved paths, helpful automation, and small changes that do not disrupt day-to-day work.

\paragraph{Internal red teaming}
Internal hack days, red-team exercises, and threat briefings again emerged as effective ways to capture leadership’s attention.
Participants noted that showing real risks in action is often what motivates executives to invest in security and support broader cultural change.

Together, these themes show that cultural change depends not just on mandates, but on making security practical and supported at all levels.

\subsection{New Ideas}

The discussion also surfaced new cultural shifts tied to developer skills and the growing use of AI.

\paragraph{Skill gaps widening with system complexity}
At this summit, participants raised concerns about a growing gap between developer skills and the increasing complexity of supply chain security.
According to them, many of their junior engineers rely on AI-assisted coding.
They often lack a basic understanding of networking, protocols, and build systems.
As a result, teams find it more challenging to explain security risks and the importance of certain practices.

\paragraph{Cultural shifts driven by GenAI adoption}
Participants noted that generative AI is transforming how developers write code, review changes, and consider risk.
They also raised concerns that increased reliance on model output may weaken review practices and introduce new cultural challenges.
Several organizations have teams that focus on shipping quickly rather than fully understanding the code they produce.
Participants noted that this can lead to software entering supply chains with less care and review.

\paragraph{Threat-intel as cultural leverage}
Participants also suggested that dedicated threat-intelligence roles can help translate attacker behavior into a sense of urgency inside organizations.
By clarifying likely risks in more precise terms, these roles might help leaders understand why action is necessary.

\section{LLMs and the Software Supply Chain}
\label{sec:llm-supply-chain}

LLMs are transforming the way people develop software.
Participants noted that this shift offers new opportunities to enhance security.
However, like a double-edged sword, it also introduces new risks to the supply chain.
Organizations now need to use AI to strengthen their existing workflows.
At the same time, they must secure the growing ecosystem of models, datasets, agents, and supporting tools that AI brings.

\subsection{Continued Themes}

Participants revisited ongoing challenges in the use and trust of LLMs in software development.

\paragraph{Developer expertise still determines AI effectiveness}
Generative AI is only helpful when developers have sufficient experience to utilize it effectively.
Participants noted that less experienced developers may follow insecure practices or misunderstand model suggestions, which can lead to errors that are difficult to detect.

\paragraph{Persistent concerns around data retention and model privacy}
Participants also raised concerns about how LLMs handle user data.
Organizations mostly rely on zero-retention policies, isolated environments, and ongoing legal reviews to address these concerns.

\paragraph{Challenges in choosing trusted models and providers}
For different organizations, it is hard to choose third-party models that they can trust.
Teams often rely on company agreements and internal checks to make these decisions.
Where data resides, whether rules are met, and legal approval also influence the final decision.

\paragraph{AI-generated noise without proper guardrails}
Participants raised concerns about LLMs producing large amounts of output that is difficult to review or of low quality.
Without robust testing, checks, and oversight, this kind of automation can end up overwhelming developers instead of helping them.

Together, these themes show that while LLMs are widely adopted, their safe and effective use still depends on developer skill, clear guardrails, and trust decisions.

\subsection{New Ideas}

The discussion also surfaced new ideas about how AI systems themselves fit into the software supply chain.

\paragraph{Emerging AI supply chain requiring provenance}
A significant new insight was treating models, datasets, fine-tuning pipelines, and agent actions as supply-chain artifacts.
Participants discussed extending SLSA and GUAC concepts to better track how AI models are built and used.
This includes recording the source of training data, maintaining hashes, and capturing inference-time provenance.

\paragraph{Identity-aware agentic workflows with tiered access}
Panelists suggested limiting high-impact agent actions, such as changing code or coordinating workflows, to more experienced developers.
Junior engineers would still be able to use agents, but only for read-only tasks.
The discussion also included ideas like giving agents clear identities.
Participants discussed defining ownership in agent-driven workflows.
They also suggested adding safeguards to prevent information from leaking between agents.

\paragraph{MCP-driven expansion of hidden supply-chain surfaces}
Another new concern this year was agents gaining access to internal systems through MCP connectors.
Some organizations mentioned they now block or tightly limit these integrations by default.
They noted that tools that can pull code from internal repositories, or even from documents, quietly expand the supply chain in ways that are hard to see and manage.

\paragraph{Privacy-preserving training for sensitive datasets}
Participants also discussed ways to train models without exposing sensitive data, such as running training inside secure enclaves.
They also discussed publishing performance attestations and preventing data leaks, noting that these measures are necessary for sensitive data, such as healthcare records.

\paragraph{Agents embedded directly into security workflows}
Another new thing this year was interest in using agentic LLMs for triage, log analysis, dependency evaluation, and automated patch recommendation. While automation has been discussed before, applying agents specifically to software supply chain tasks is a new direction.

\section{Executive Summary}
\label{sec:summary}

The September 2025 S3C2 Secure Software Supply Chain Summit brought together practitioners from multiple organizations to share firsthand experiences and lessons learned in defending and improving modern software supply chains. Six panels captured how teams are adapting to a rapidly changing and increasingly complex ecosystem: vulnerable dependencies (Sec.~\ref{sec:vuln-dependencies}), component and container choice (Sec.~\ref{sec:choices}), malicious commits (Sec.~\ref{sec:malicious-commits}), build infrastructure (Sec.~\ref{sec:build-infrastructure}), security culture (Sec.~\ref{sec:culture}), and the role of large language models (LLMs) in the software supply chain (Sec.~\ref{sec:llm-supply-chain}). While many discussions revisited challenges raised in earlier summits, participants also shared new details, perspectives, and real-world examples.

In the \emph{Vulnerable Dependencies} panel, participants highlighted fatigue from excessive alerts from the SCA tools. They discouraged thinking in terms of CVE only.
For getting to the root of vulnerabilities, participants emphasized the need for richer context, including maintainer trust, reachability, and lifecycle signals.
New ideas focused on pushing dependency risk signals directly in developer workflows and treating IDEs as part of the supply chain.
Some participants also pointed out that shifting some trust decisions upstream through vendor-curated library sets would greatly benefit small companies.

The \emph{Component and Container Choice} session highlighted the lack of standardized methods for evaluating dependencies and container images.
Participants also emphasized the need to integrate scanning, registries, and policy checks more closely with developer workflows.
They emphasized that these controls need to add as little friction as possible to the developer workflow.

The \emph{Malicious Commits} panel focused on the difficulty of detecting harmful changes in shared codebases. Participants noted that malicious commits can easily be mistaken for honest mistakes in the codebase. The discussion also introduced interesting signals, such as changes in contributor behavior and capability levels, such as a contributor pushing a change outside their regular work hours.

In the \emph{Build Infrastructure} panel, workload identity emerged as the central theme. Participants noted that securing CI/CD systems increasingly depends on representing, authenticating, and verifying the actors and agents performing builds.

In the \emph{Culture} session, teams shared strategies that make security easier. Developer-first approaches, such as paved paths, helpful automation, and small changes that do not disrupt day-to-day workflows, encourage adoption of security practices. Participants also suggested that dedicated threat-intelligence roles can help turn attacker behavior into a clearer sense of urgency. By explaining likely risks in plain terms, these roles can help leaders understand why action is needed.

Finally, the \emph{LLMs and Supply Chain} panel examined how generative AI is reshaping both sides of the problem: AI as a tool for security triage and as a new supply chain that requires its own provenance, isolation, and governance. The discussions mainly centered on the supply chain for AI and the role of AI within the supply chain.

\section{Acknowledgements}
We sincerely thank all Summit participants for their time, openness, and thoughtful contributions. Their shared experiences and suggestions greatly enriched the discussion. The Summit was organized and recorded by Dr. Laurie Williams, Dr. Dominik Wermke, and Md Atiqur Rahman.
This material is based upon work supported by the National Science Foundation Grant Nos. 2207008, 2206859, 2206865, and 2206921.
These grants support the Secure Software Supply Chain Summit (S3C2), consisting of researchers at North Carolina State University, Carnegie Mellon University, University of Maryland, and George Washington University. 
Any opinions expressed in this material are those of the author(s) and do not necessarily reflect the views of the National Science Foundation.

\bibliographystyle{ACM-Reference-Format}
\bibliography{literature}

\appendix

\section{Initial Discussion Questions}
\label{questions}
\begin{enumerate}

\item \textbf{Vulnerable Dependencies.} What processes and/or tools do you use to find out that you have a vulnerable dependency? What is your process for evaluating/prioritizing what dependencies to update and actually updating vulnerable dependencies?   Do you push a new dependency version with a major or minor release? How do you deal with what happened with NVD last year? Are you using OSV or GHSA more? Do you worry about incorrect and missing information in vulnerability databases?

\item \textbf{Component and Container Choice.} What is the process for bringing a new component or container into a product?  Do you use OpenSSF Scorecard or other metrics to help you with your decision-making?  Are component choices re-evaluated periodically?

\item \textbf{Malicious Commits.}  How can malicious commits be detected? What do you think signals a suspicious/malicious commit?  What role does the ecosystem play in detecting malicious commits?

\item \textbf{Build Infrastructure.}  What is being done (or should be being done) to secure the build and deploy process/tooling pipeline (a.k.a SLSA practices)?  Are you working toward reproducible builds?  Do you run your own build server or cloud services?  Do those who use GitHub actions use self-runners?  Are you seeing increased attention by security researchers and attackers on weaknesses in GitHub Actions?

\item \textbf{Culture.} What changes have you made to support supply chain security/executive order compliance?  What do you think is needed for nurturing such a security-benefiting culture?

\item \textbf{LLMs and Supply Chain.} How are you leveraging the recent advances in ML/AI in securing your software supply chain?  What are the policies around the use of code generation tools in your company?  Do you have a process for choosing safe/secure third-party AI models?

\end{enumerate}

\end{document}